# Reduction in Packet Delay Through the use of Common Buffer over Distributed Buffer in the Routing Node of NOC Architecture

*Nilesh A. Mohota #1, Sanjay L. Badjate #2*
*#1Department of Electronics Engineering, Rashtrasant Tukdoji Maharaj Nagpur University Nagpur, India*
*#2Department of Electronics & Telecommunication Engineering, S. B. Jain Institute of Technology, Management and Research, Nagpur, India*
[1] nileshmohota@gmail.com

***Abstract*— The continuous innovation of semiconductor technology enables more complex System on-Chip (SoC) designs. Tens, even hundreds of intellectual properties (IPs) are integrated into an SoC to provide various functions, including communications, networking, multimedia, storage, etc. The bus scheme connects multiple IP cores with a cost efficient shared medium. The bus-based scheme still fails to satisfy the requirements of future SoC mainly due to two major drawbacks. Non-scalable and the bandwidth is shared by all IPs and thus the bus becomes the performance bottleneck when more and more IPs are connected. In order to interconnect such a high number of elements on a die, researchers have turned to Network On Chip as a replacement to conventional shared buses and ad-hoc wiring solutions. They are attractive due to their regularity and modular design, which can lead to better routability, electrical characteristics and fault tolerance.**

**Performance evaluation of the routing node in terms of latency is the characteristics of an efficient design of Buffer in input module. It is intended to study and quantify the behavior of the single packet array design in relation to the multiple packet array design. The utilization efficiency of the packet buffer array improves when a common buffer is used instead of individual buffers in each input port.**

**First Poisson's Queuing model was prepared to manifest the differences in packet delays. The queuing model can be classified as (M/M/1); (32;FIFO). Arrival rate has been assumed to be Poisson distributed with a mean arrival rate ($\lambda$) of 10 x $10^6$. The service rate is assumed to be exponentially distributed with a mean service rate of 10.05 x $10^6$. It has been observed that latency in Common Buffer improved by 46% over its distributed buffer.**

**A Simulink model later simulated on MATLAB to calculate the improvement in packet delay. It has been observed that the delay improved by approximately 40% through the use of a common buffer. A verilog RTL for both common and shared buffer has been prepared and later synthesized using Design Compiler of SYNOPSYS. In distributed buffer, arrival of data packet could be delayed by 2 or 4 clock cycles which lead to latency improvement either by 17 % or 34 % in a common buffer**

Keywords: **A**rrival rate ($\lambda$), service rate, FIFO, latency, Simulink model, packet array, IP mapping.

## I. INTRODUCTION

A packet-switched 2-D mesh is the most used and studied topology so far. It is also a sort of an average NOC currently. Good results and interesting proposals provoke design engineers to use this topology as the base [1] .The key research problems in the design of NOCs include but are not limited to topology, channel width, buffer size, floor plan, routing, switching, scheduling, and IP mapping [2]. Additionally, [3] lists research issues to be application modeling and optimization, NOC communication architecture analysis and optimization, NOC communication architecture evaluation, and NOC design validation and synthesis. The most important metrics for NOCs are application runtime, silicon area, power consumption, latency and throughput. All these are to be minimized and usually appropriate trade-off is sought [4]. The required silicon area is the most commonly reported value (77%) followed by latency (55%) and maximum operating frequency (50%). The other metrics have lower occurrence [1].

In this regard, the current work is related to optimization of buffers in the router design so as to achieve lower latency. The router consists of four major components: Crossbar, Switch, FIFO & Buffers, see fig.

1.1. The no. of buffers in the routing node are dependent on the type of topology and the no of adjacent nodes [1]. In 2D Mesh NoC topology, a node may have four adjacent contemporary nodes. In the existing work [3], the routing node holds four buffers each to accommodate input stream of data from adjacent node. The buffers are also referred as packet arrays hold the stream of the incoming packets and dispatch them once scheduling is done.

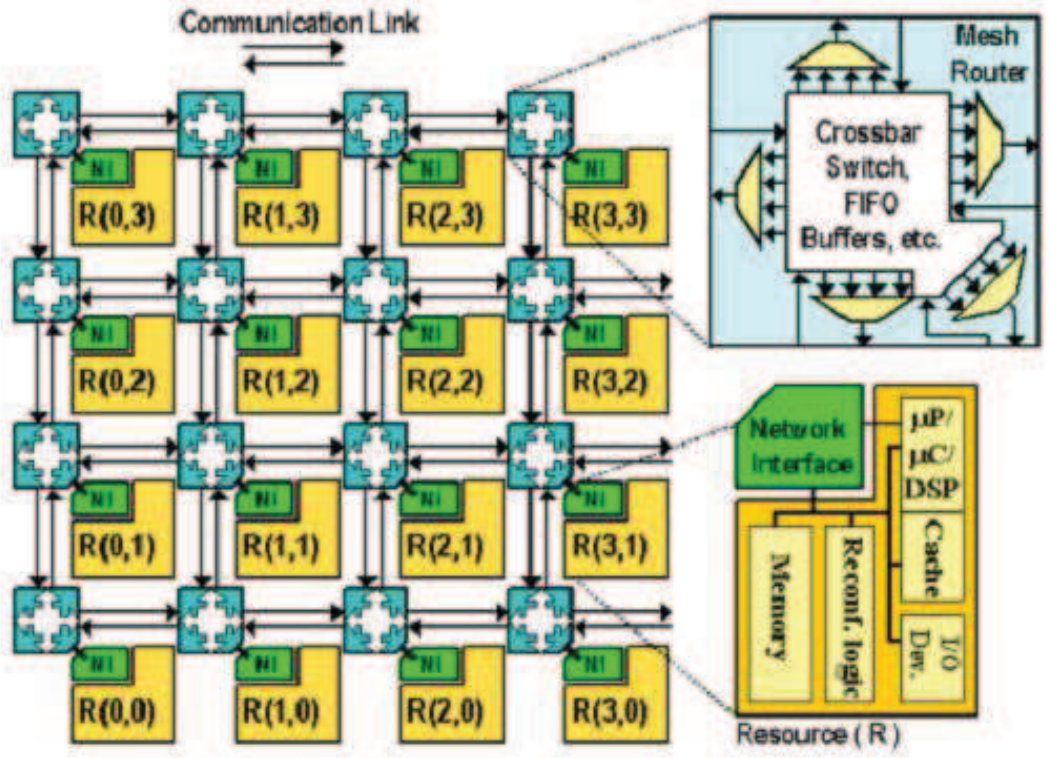

Fig. 1.1 A generalized 2D Mesh NoC

## II. OVERVIEW OF NETWORK ON CHIP ARCHITECTURE

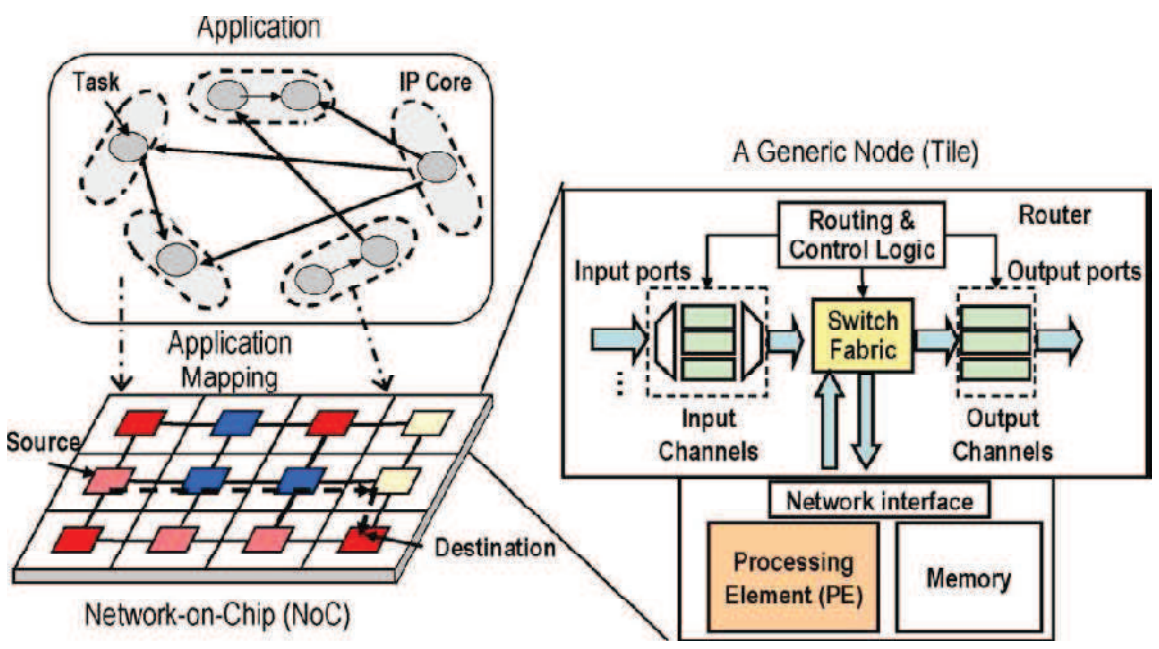

Fig. 2.1 Generic NOC structure

A NOC consists of routers/switch, links, and network interfaces (Fig. 2.1). Routers direct data over several links (hops). Routers further consists of a scheduler, buffer to store the incoming data packet and the crossbar. Topology defines their logical lay-out (connections) whereas floor plan defines the physical layout. The function of a network interface (adapter) is to decouple computation (the resources) from communication (the network). Routing decides the path taken from source to the destination whereas switching and flow control policies define the timing of transfers **[2][5].** Task scheduling refers to the order in which the application tasks are executed and task mapping defines which processing element (PE) executes certain task. IP mapping, on the other hand, defines how PEs and other resources are connected to the NoC.

### A. *Router Structure*

NOC architectures are based on packet-switched networks, see figure 2.1.a. This has led to new and efficient principles for design of routers for NOC **[6].** Assume that a router for the mesh topology has four inputs and four outputs from/to other routers, and another input and output from/to the Network Interface (NI). Routers can implement various functionalities - from simple switching to intelligent routing. Since embedded systems are constrained in area and power consumption, but still need high data rates, routers must be designed with hardware usage in mind. For circuit-switched networks, routers may be designed with no queuing (buffering). For packet-switched networks, some amount of buffering is needed, to support bursty data transfers [3].

Buffers can be provided at the input, at the output, or at both input and output [7].Various designs and implementations of router architectures based on different routing strategies have been proposed in the literature. Wolkotte et al. proposed a circuit switched router architecture for NOC [8], while Dally and Towles proposed a packet switched router architecture [9]. Albenes and Frederico provided a wormhole-based packet forwarding design for a NOC switch [10].

In this paper, the buffers in the design of the routers are based on the principle of virtual output queuing since it is simple and reduces the risk of Head of Line Blocking [11] [12] [13]. In this paper, the scheduling policy embodied in the router is based on Iterative SLIP algorithm. iSLIP uses round-robin to choose on port among those contending. This permits simpler hardware implementations compared, besides making iSLIP faster. iSLIP achieves close to maximal matches after just one or two iterations.. iSLIP achieves 100% throughput under uniform traffic and the round robin policy ensures fairness among contenders. Even though its behavior may be unstable under bursty traffic, iSLIP is commonly implemented in commercial switches due to its simplicity [14]. This algorithm becomes more silicon area efficient if it is implemented with its folding concept [15].

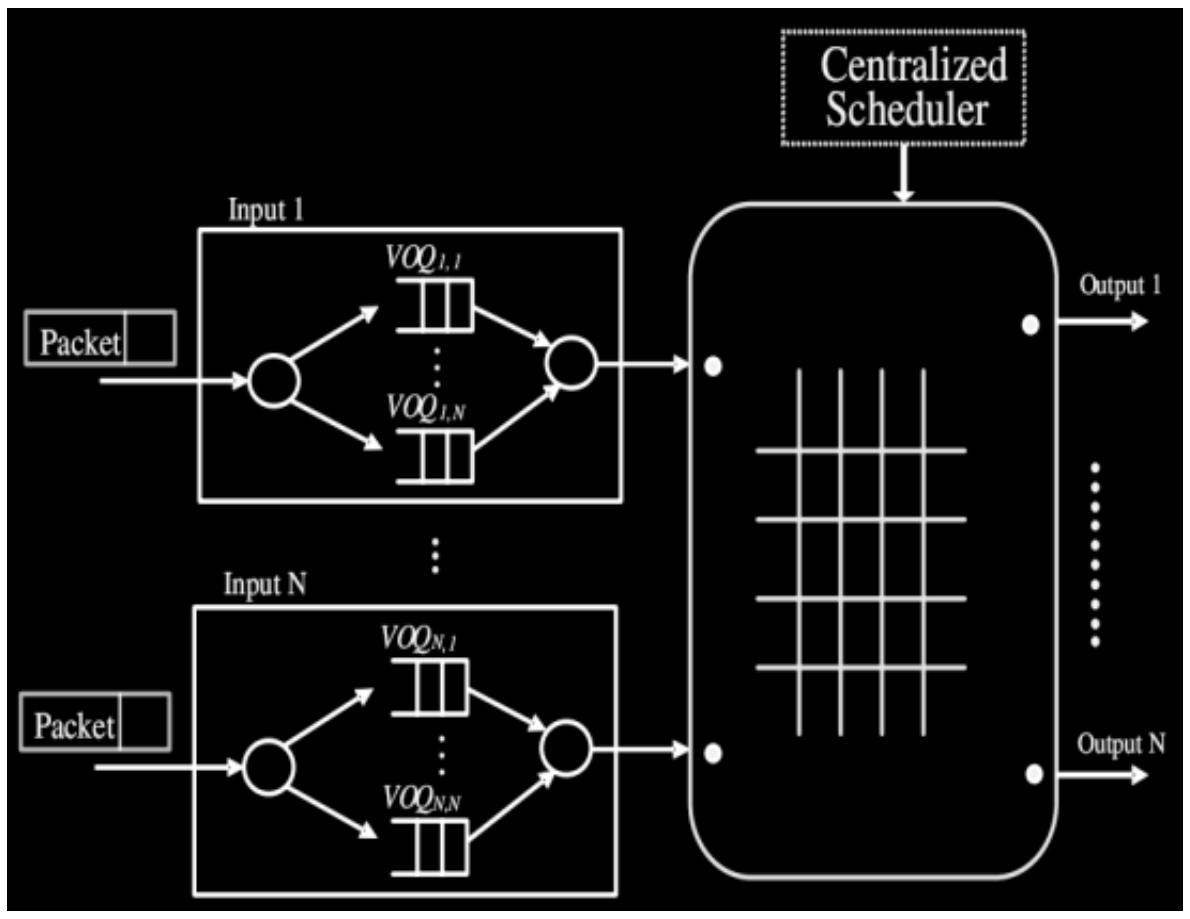

Fig 2.1.a NOC Router Components

## III. RELATED WORK

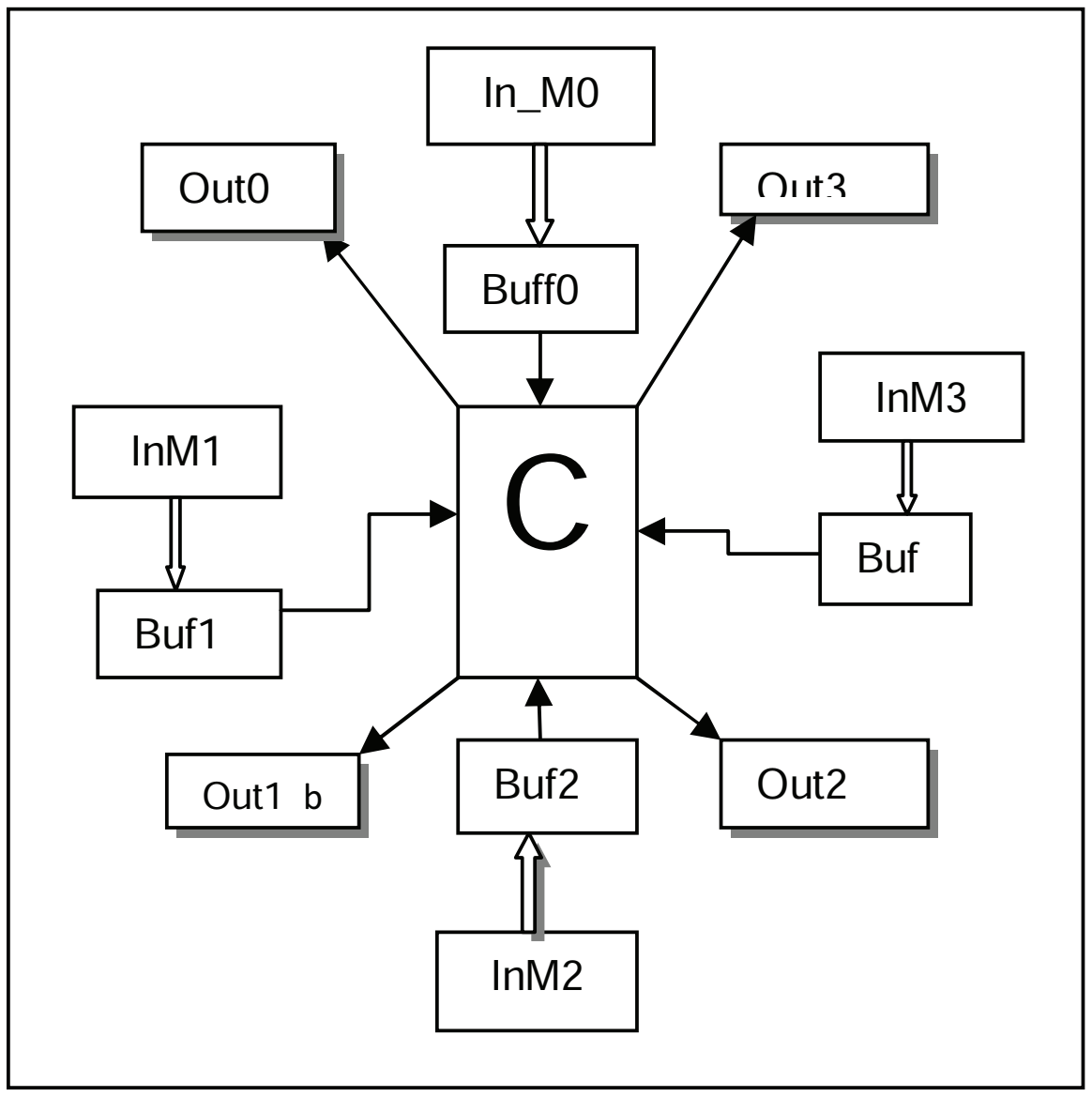

Fig. 3.1 Each input port has its own buffer

The current design of the input block is based on virtual output queues. The queues are maintained in a packet array. Queue size is dynamically determined depending on the arrival pattern of the data. If more data is destined for output port "m", then correspondingly, more buffer space, and hence, a longer queue is maintained for data packets to be routed to output port "m," subject to the maximum space available in the packet array (Fig.3.1).

An alternative design is based on using a common packet for all the input ports. For example, if the crossbar switch consists of four input ports, then the original design calls for four packet arrays. The proposed design would utilize one common packet array for all the four input ports (Fig.3.2).

It is intended to study and quantify the behavior of the single packet array design in relation to the multiple packet array design. Intuitively, a common packet buffer would result in better utilization of available buffer space. This in turn would translate into lower delays in transmission.

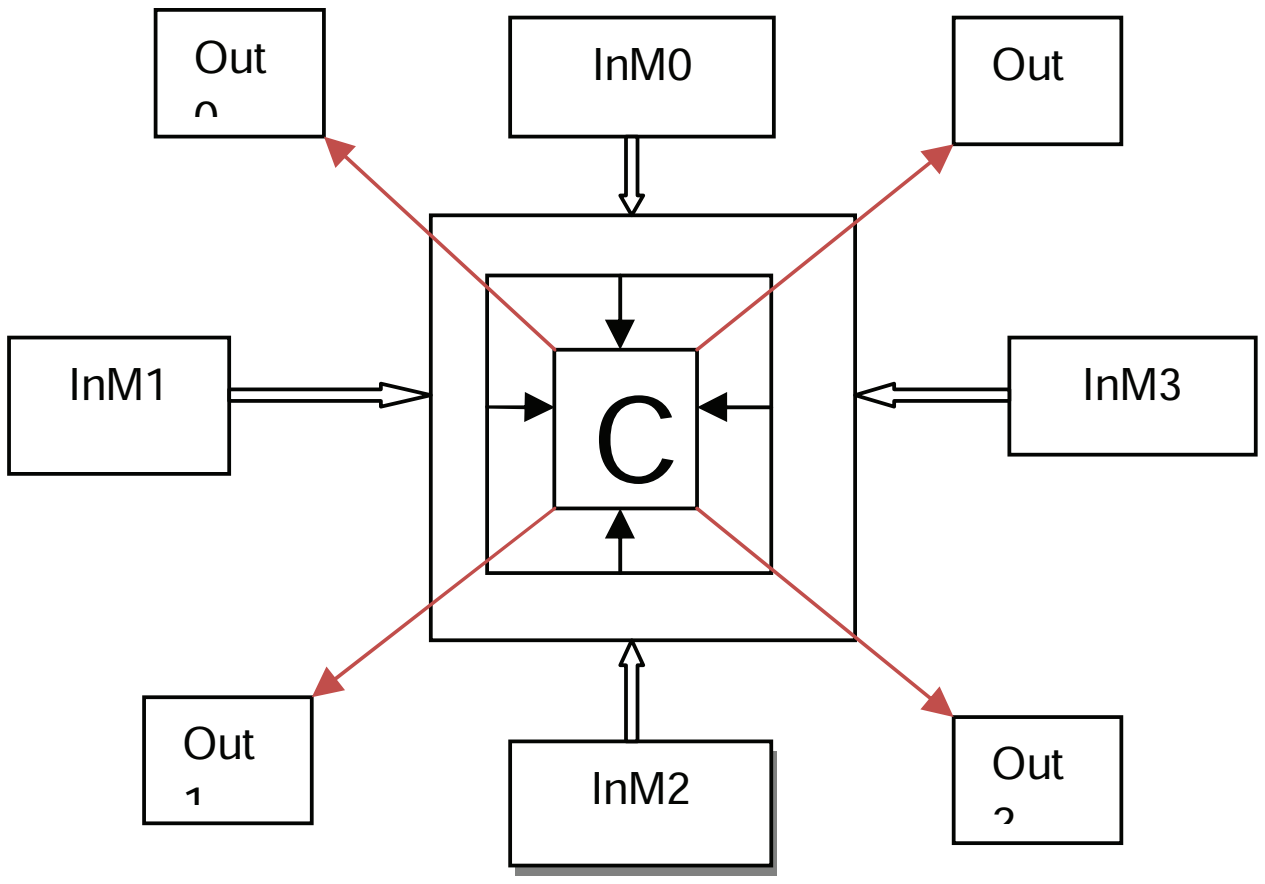

Fig. 3.2 Common buffer shared between all input ports

## IV. POISSON'S QUEUING THEORY [18]

- Mean Arrival Rate (λ) = $10 \times 10^6$
- Mean Service Rate (μ) = $10.05 \times 10^6$
- Traffic Intensity (ρ) = λ / μ = 0.995
- The queuing model is classified as: (M/M/1); (32;FIFO).
- In the first model there is a common buffer having capacity of 128 packets, see fig. 5.1
- In the second model there are 4 independent queues each having capacity of 32 packets, see fig. 5.2
- Latency = E(n) / λ

### A. *Effect of Having a Single Input Buffer*

Some assumptions have been made regarding arrival rate and service rate of packets. Arrival has been assumed to be Poisson distributed with a mean arrival rate ($\lambda$) of 10 x $10^{6.}$ The service rate is assumed to be exponentially distributed with a mean service rate of 10.05 x $10^{6.}$ Traffic intensity • = • / • = 0.995.

In the first model there are four independent queues each having capacity of 32 packets. The queuing model can be classified as (M/M/1); (32;FIFO) [18]. The system stops taking further input when queue size reaches 32.

The average no. of packets in the system are given by

$$P_n = \frac{(1-\rho).\rho^n}{1-\rho^{N+1}}$$

$$P_{32} = \frac{(1-0.0995)\times 0.995^{32}}{1-0.995^{33}}$$

$$P_{32} = \frac{0.005\times 0.8151}{1-0.8475} = 0.027$$

i.e. 2.7%

*1) Latency for N = 32:* Avg. no. of packets in system

$$E(n) = \frac{\rho[1-(N+1)\rho^N + N\rho^{N+1}]}{(1-\rho)(1-\rho^{N+1})}$$

$$= \frac{0.995\times(1-33\times 0.995^{32} + 32\times 0.995^{33})}{(1-0.995)(1-0.995^{33})}$$

$$= \frac{0.995(1-28.1+27.12)}{0.005\times 0.152}$$

$$= 2618$$

$$\text{Latency} = \frac{2618}{10\times 10^6} = 2618\times 10^{-7}$$

*2) Latency for N = 128:* In this case four input queues are merged into a single input queues of capacity 128. Since all the sources put data into the same queues, arrival rate is assumed to be 4 times higher. The servicing rate is also assumed to be four times higher. Traffic intensity remains same at 0.995.

$$E(n) = \frac{0.995(1-129\times 0.995^{128} + 128\times 0.995^{129})}{(1-0.995)(1-0.995^{129})}$$

$$= \frac{0.995(0.136)}{0.005\times 0.476} = 56$$

$$\text{Latency} = \frac{56}{4\times 10\times 10^6} = 14\times 10^{-7}$$

It has been seen theoretically that latency reduces merging of the queues.

A MATLAB model used to show quantitatively how performance is improved in a common packet array design.

## V. SINGLE BUFFER OF SIZE 128 PACKETS VERSUS 4 BUFFERS OF SIZE 32 PACKETS

The block diagram of the Simulink model is given below.

Single 128 packet buffer with 4x4 Scheduler fig 5.1.

Four 32 packet buffers with 4x4 Scheduler fig 5.2.

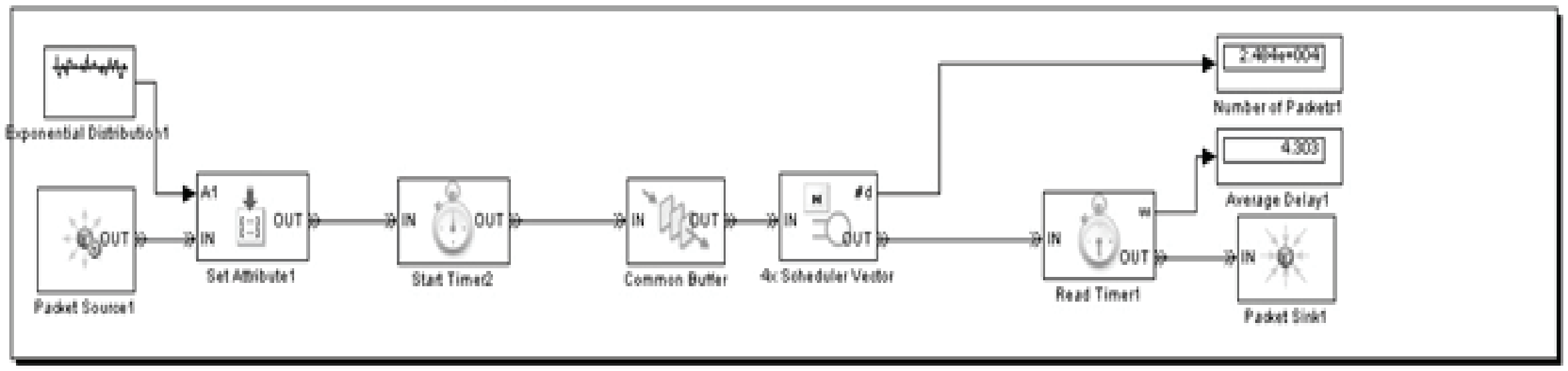

Fig. 5.1 Simulink model for single buffer

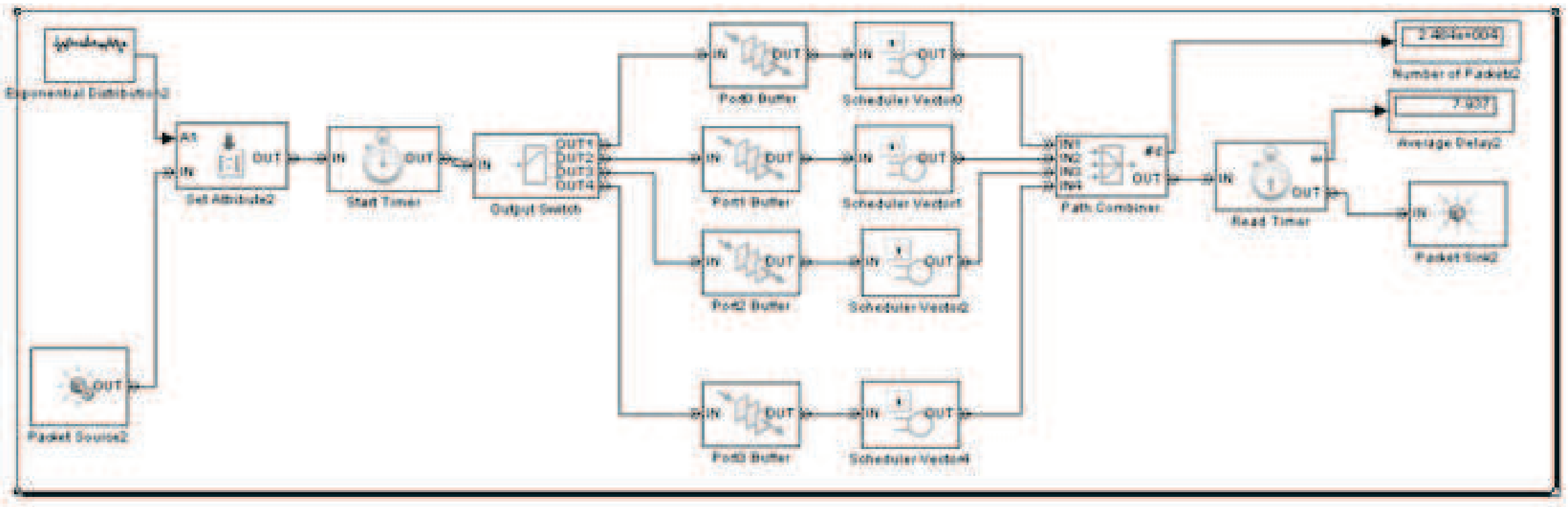

Fig. 5.2 Simulink model for 4-buffers

The first block is labeled "Exponential Distribution1." This block specifies packet arrival time. The packet arrival pattern is an exponential distribution. The block labeled "Packet Source1" generates packet events. The "Set Attribute1" block combines the effects of "Exponentia Distribution1" with "Packet Source1" to generate packet entities at time intervals specified by the exponential distribution mentioned above. The arrival of a packet event at the "Start Timer2" block causes the simulation timer to start. Generated packets are stored in a common 128 packet buffer designated "Common Buffer." The packets leave the common buffer when they are serviced by the scheduler vector. The scheduler vector is generated for four input ports by "4x Scheduler Vector." The total number of packets served are recorded in the "Number Serve1" block. Whenever a packet leaves the buffer, the departure time is recorded by the "Read Timer." The packet exits the simulation flow through the "Entity Sink1" block. The average time spent by the packet in the buffer is captured by the "Average Delay1" block.

The behavior of the dedicated 32 packet buffer model differs only in two components, "Output Switch" and "Path Combiner." The "Output Switch" block demultiplexes the generated packets into their respective input port packet buffers. The "Path Combiner" aggregates the output stream to help calculate total number of packets served and average time spent by packet waiting for service. The simulation was run for 50000 packets. The packet generation rates for both models are identical, using the exponential distribution for inter-arrival times. Multiple simulation runs were performed to verify the average delays observed.

## VI. RESULTS

TABLE I
LATENCY IN COMMON BUFFER VS DISTRIBUTED BUFFER

| **SIMULATION MODEL** | **COMMON BUFFER (128 PACKETS)** | | **DISTRIBUTED BUFFERS (4 x 32 PACKETS)** | |
|---|---|---|---|---|
| Average Latency | 1.4 time unit (T) | 4.3 time unit (M) | 2.6 time unit (T) | 7.9 time unit (M) |
| | | | | |

T - Latency derived using Poisson's Queuing Theorem.

M – Latency derived using SIMULINK MATLAB MODEL

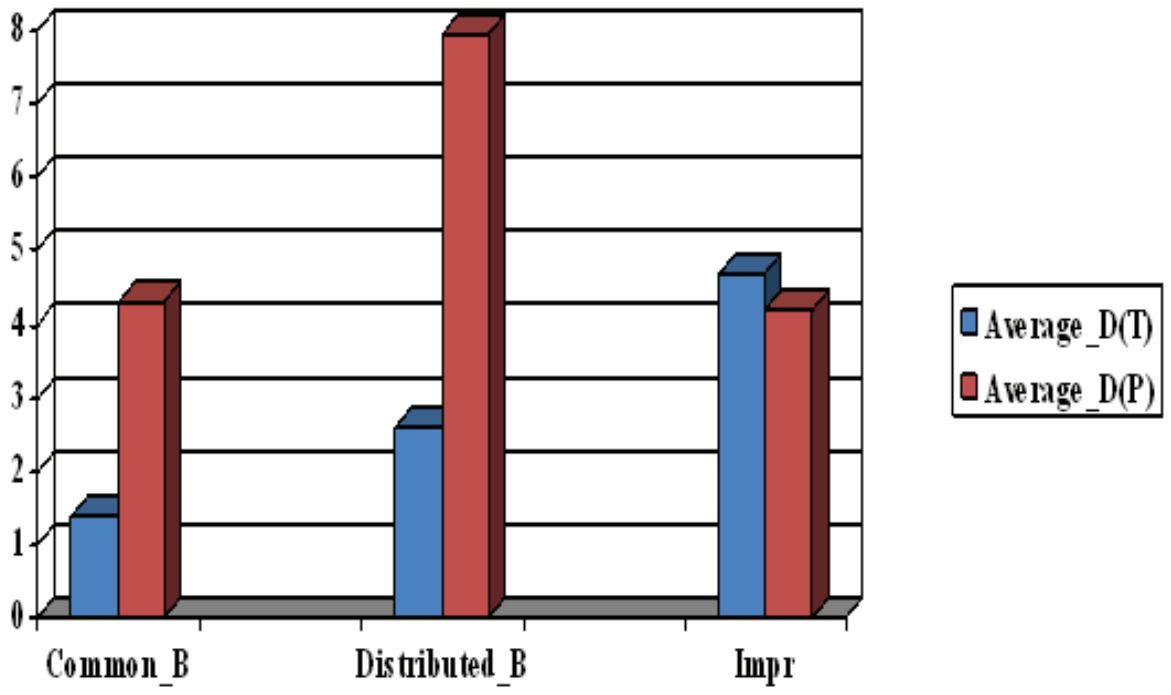

46% Theoretical,

40% Practical.

A. *Latency in HDL Model*

MIN LATENCY – It is defined as the total amount of time from the start of packet transmission at source to start of packet reception at destination.

TABLE II
**LATENCY IN COMMON BUFFER VS DISTRIBUTED BUFFER HDL MODEL**

| **SIMULATION MODEL** | **COMMON BUFFER (128 PACKETS)** | **DISTRIBUTED BUFFERS (4 x 32 PACKETS)** |
|---|---|---|
| Average Latency | [2+4+4] = 10 CCs x 4 ns = 40 ns | [2+4+4+2] = 12 CC x 4 ns = 48 ns |

**Common Buffer:**

2 CCs – to store the packets in two phases in packet array

4 CCs – to reach to scheduling decision

4 CCs – to travel to destination

**Distributed Buffer:**

In normal case, Avg. latency may be 10 CCs for data packet to move to its destination as described in Common Buffer. However, it may take additional 2 or 4 CCs if the desired packet array is hugely crowded. Synthesis using Design Compiler of SYNOPSY has been done and it has been observed that the longest combinational path of 4 ns i. e. clock period was compatible to both the design.

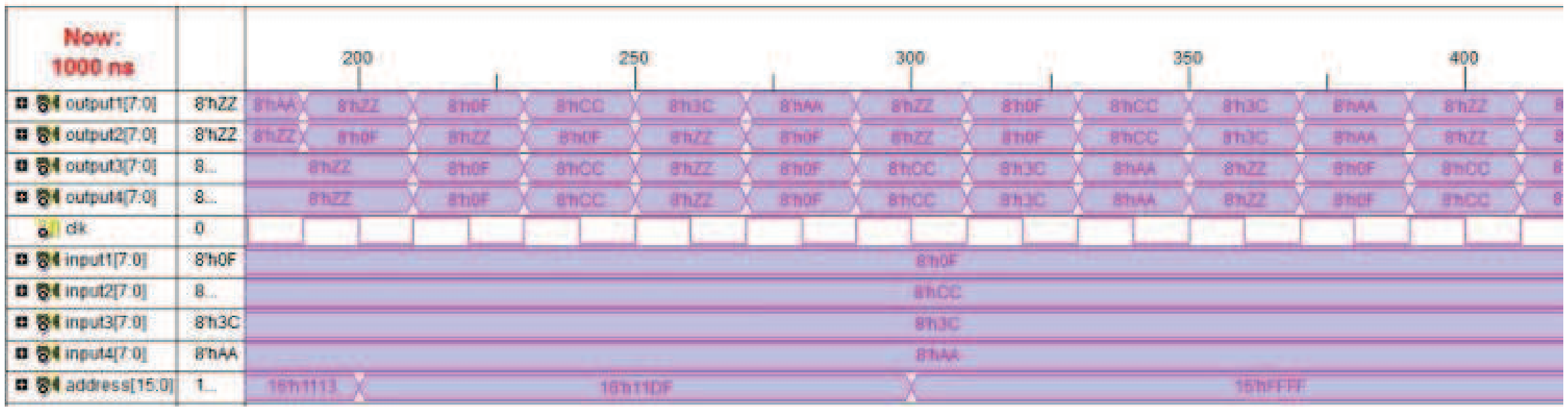

Fig 6.1: A 4x4 Routing node

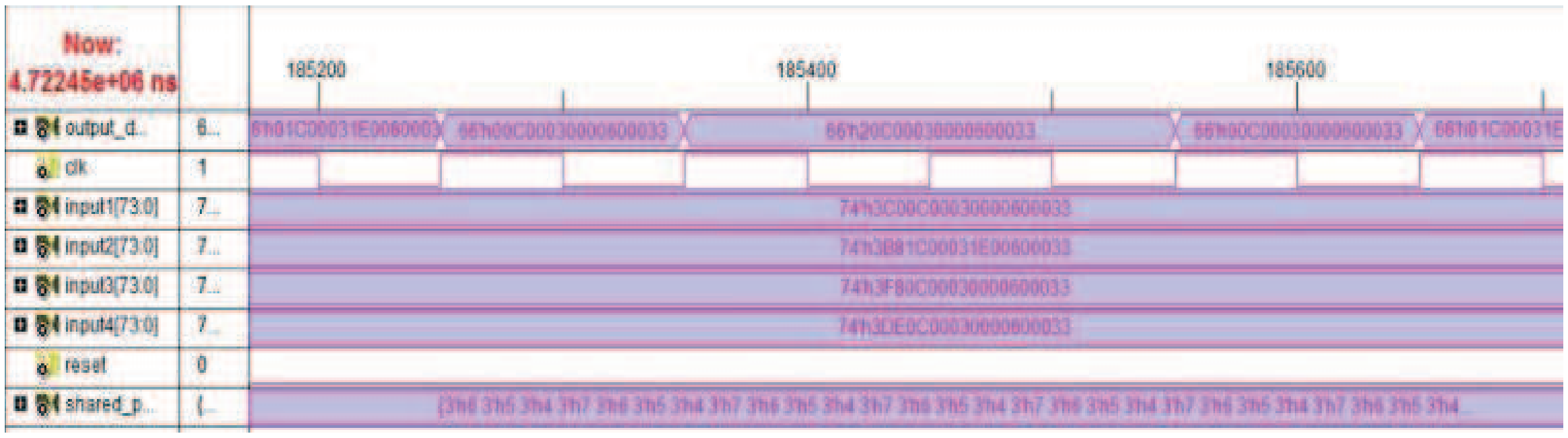

Fig 6.2: Common Buffer waveform

## VII. CONCLUSION

The utilization of buffer memory space is improved by replacing separate buffers in each input block with a shared buffer common to all input blocks. This is manifested in the form of lower delay in transferring a packet from the input to the output. First Poisson's Queuing model was prepared to manifest the difference in latencies. It has been observed that latency in Common Buffer improved by 46% over its distributed buffer.

A Simulink model later simulated on MATLAB to calculate the improvement in packet delay. It has been observed that the delay improved by approximately 40% through the use of a common buffer.

A verilog RTL for both common and shared buffer has been prepared and later synthesized using Design Compiler of SYNOPSYS. In distributed buffer, arrival of data packet could be delayed by 2 or 4 clock cycles under heavy and undistributed traffic. Under such circumstances, latency improvement could be claimed either by 17 % or 34 % in a common buffer.